\documentclass[draft]{agujournal2019}
\usepackage{url} 
\usepackage[inline]{trackchanges} 
\usepackage{soul}
\draftfalse

\journalname{JGR: Space Physics}

\begin{document}

%
%


\title{Ionospheric Response to the May 11, 2024, Geomagnetic Superstorm over Ecuador}

%
%




\authors{Ericson Lopez\affil{1,2}, Hugo Barbier\affil{1}, William Carvajal\affil{2}, Leonel Guamán\affil{2}}


\affiliation{1}{Escuela Politécnica Nacional, Facultad de Ciencias, Departamento de Física, Quito, Ecuador}
\affiliation{2}{Escuela Politécnica Nacional, Observatorio Astronómico de Quito/ Observatorio Nacional, Quito, Ecuador}




\correspondingauthor{E. D. López}{ericsson.lopez@epn.edu.ec}



\begin{keypoints}
\item Impact of the geomagnetic storm: A G5 geomagnetic storm (10-13 May 2024) caused an unusual drop in Total Electron Content (TEC) above Galápagos, Ecuador.

\item Unusual TEC behavior: The drop in TEC during the storm peak was likely due to rapid atmospheric changes, followed by a gradual increase in TEC after the storm.

\item Importance of Continuous Monitoring: The study highlights the need for continuous monitoring to understand the effects of geomagnetic storms on satellite communications and navigation systems.
\end{keypoints}

%
%

%
%


\begin{abstract}

This study investigates the impact of the G5 geomagnetic storm on Total Electron Content (TEC) derived from the Global Positioning System (GPS) in Galápagos, Ecuador (geographic latitude 0.1807° S, longitude 78.4678° W) during May 10-13, 2024. Using vertical TEC (VTEC) data from a single pseudorandom noise (PRN) code, along with the average VTEC from the same PRN collected over the ten days before the storm, referred to as background TEC, to analyze the variations in TEC.

Our findings indicate that during the main phase of the storm on May 10-11, 2024, TEC experienced a notable decrease, which contrasts with the typical responses observed in previous storms. This decrease can be attributed to rapid recombination processes and potential plasma instabilities triggered by the storm. In the recovery phase following the main storm, a gradual increase in TEC was observed, illustrating the complex dynamics of the ionosphere in response to geomagnetic disturbances.

This study underscores the variability in TEC responses during geomagnetic storms. It highlights the importance of real-time monitoring to improve our understanding of the implications for satellite communication and navigation systems.

\end{abstract}

\section*{Plain Language Summary}
 In this study, we examined how a strong geomagnetic storm (G5 level) affected the Total Electron Content (TEC) in the atmosphere above Galápagos, Ecuador, from 10 to 13 May 2024. TEC measures the number of electrons in a certain area of the ionosphere that can affect satellite signals. We used data from the Global Positioning System (GPS) to analyze changes in TEC.

Our results showed that during the peak of the storm on 10-11 May 2024, the TEC dropped significantly, which is different from what we usually see in other storms. This drop was likely caused by fast changes in the atmosphere and possible instabilities caused by the storm. After the storm, the TEC gradually increased, showing how the ionosphere can change in complex ways during and after a geomagnetic disturbance.

This study highlights that TEC can respond in different ways during geomagnetic storms and emphasizes the need for continuous monitoring to better understand how these storms affect satellite communications and navigation systems.

%
%

%


%
%
%
%

\section{Introduction}
From 10-13 May 2024, an exceptionally strong geomagnetic storm, classified as G5, occurred at the peak of solar cycle 25. This event was significant due to its impact, which resulted from multiple observed coronal mass ejections (CMEs) originating from the Sun \cite{kwak2024}. 

Coronal mass ejections (CMEs) are massive expulsions of plasma and magnetic fields from the Sun’s corona into space, capable of significantly influencing Earth’s space weather \cite{shen2022}. These events can cause geomagnetic storms that affect telecommunications, navigation systems, and power grids \cite{pulkkinen2017}.

Geomagnetic storms arise from the interaction of the solar wind with the magnetic field of the Earth \cite{sun2024}. These storms can cause large-scale disturbances in the ionosphere, including irregularities and rapid fluctuations in electron content. The resulting variations in vertical total electron content (VTEC) can pose challenges to accurate modeling and prediction \cite{chen2024}.

Recently, GPS measurements have been utilized to study the effects of geomagnetic storms on total electron content (TEC). Research focused on regions affected by the Equatorial Ionization Anomaly (EIA) is particularly important, as the equatorial plasma fountain is highly sensitive to electrical disturbances \cite{shagimuratov2024}.

The EIA is mainly associated with the distribution of electron density in the ionosphere rather than directly with the magnetic field of the Earth \cite{abdu2005}. However, its presence and behavior are closely linked to the magnetic field, especially its horizontal component at the magnetic equator. The near-horizontal magnetic field lines at the equator are crucial for the formation of the EIA through the equatorial fountain effect \cite{balan2018}. During the day, solar heating generates an eastward electric field at the equator. This electric field interacts with the Earth’s magnetic field to produce an upward drift (E × B). Ionized particles are lifted to higher altitudes by this drift and subsequently move along the magnetic field lines to higher latitudes, creating two crests of high electron density. Enhanced electric fields during storms can intensify this E × B drift, leading to a more pronounced EIA \cite{aa2024}.

Geomagnetic storms induce fluctuations in Earth’s magnetic field, particularly at high latitudes \cite{kamide1998}. These fluctuations can propagate to lower latitudes, including the equatorial region. Magnetic field disturbances can induce electric fields and currents in the ionosphere, known as disturbance dynamo effects, which can modify existing ionospheric electric fields \cite{tahir2024}. During the initial phase of a geomagnetic storm, high-latitude electric fields can penetrate to equatorial latitudes almost instantaneously. These prompt penetration electric fields can enhance the eastward electric field at the equator, increasing the upward E × B drift and intensifying the EIA.

Geomagnetic storms also cause variations in the Earth’s magnetic field, known as magnetic field perturbations, which are recorded as changes in both horizontal and vertical components \cite{hayakawa2021}. These variations can be observed globally, with more pronounced effects at high and equatorial latitudes as a result of intensified auroral and equatorial currents.

The Equatorial Electrojet (EEJ) typically increases during the daytime; enhanced eastward electric fields boost the EEJ, resulting in a higher magnetic field at ground level in the equatorial region \cite{yamazaki2017}. Conversely, if the electric fields weaken or reverse (for example, during nighttime or due to disturbance dynamo effects), the EEJ diminishes, leading to a decrease in the magnetic field. Enhanced electric fields and intensified ionospheric currents, such as auroral and equatorial electrojets, generally lead to an increase in the magnetic field, particularly noticeable during the main phase of a geomagnetic storm at high and equatorial latitudes. Opposing electric fields, disturbances in ionospheric currents, and the recovery phase of geomagnetic storms can result in a decrease in the magnetic field \cite{li2024}.

Factors such as geomagnetic storms, solar flares, and other space weather events can induce sudden and severe disturbances in the ionosphere, complicating the generation of accurate VTEC maps during these periods \cite{chen2024}.

Space weather, primarily driven by solar activity, significantly affects the accuracy of VTEC (Vertical Total Electron Content) maps by introducing variations and disturbances in the Earth’s ionosphere \cite{luo2023}. Key ways in which space weather impacts the accuracy of VTEC maps include increased ionospheric variability and disruptions caused by solar phenomena.

The Dst and Kp indexes are important tools for monitoring and understanding geomagnetic activity \cite{matzka2021}. These indices track the strength of magnetic storms triggered by solar activity, such as solar flares and CMEs, which can affect the Earth’s magnetosphere and ionosphere. The Dst index measures the globally averaged strength of geomagnetic disturbances, quantifying how much Earth’s magnetic field is disturbed during these events \cite{lin2021}. It typically ranges from negative values (indicating geomagnetic storm conditions) to positive values (indicating quieter conditions), usually falling between -20 and +20 nanotesla (nT). The Dst index is primarily used to assess the severity and duration of geomagnetic storms, which can impact technological systems such as satellites, power grids, and radio communications.
The Kp index measures the level of geomagnetic activity on a global scale, providing a real-time assessment of storm intensity \cite{matzka2021}. It ranges from 0 to 9, with increasing values indicating higher levels of geomagnetic activity. The Kp index is vital for space weather forecasting, helping predict the impacts of geomagnetic storms on Earth’s magnetosphere and technological systems.
Both indices are crucial for monitoring and predicting space weather, as they provide quantitative measures of geomagnetic disturbances.

\vspace{0.3 cm}
In this study, we examined the effects of geomagnetic storms on GPS-derived VTEC at the Galapagos receiver GLPS (geomagnetic latitude $0^{\circ} 44^{'} 34.97^{''}S$ , geomagnetic longitude $90^{\circ} 18^{'}13,2^{''}W$), situated below the crest region of the Equatorial Ionization Anomaly (EIA). Our analysis aimed to correlate the filtered CME data with the geomagnetic indices Dst and Kp, focusing on the impact of the severe geomagnetic storm on 11 May 2024, on variations in GPS-derived total electron content (TEC) in Galápagos, Ecuador. Our results indicated that during the G5 geomagnetic storm of May 2024, vertical TEC (VTEC) values decreased during the main phase of the storm.
 
\section{Procedure to Obtain TEC from GPS Receivers}
The ionosphere exhibits a refractive index at radio frequencies that deviates from unity, affecting GPS signals as they travel from the satellite to the ground receiver \cite{misra2006}. A major consequence of this interaction is the additional delay that GPS signals experience while traversing the ionosphere, which is directly proportional to the total electron content (TEC). TEC is defined as the total number of free electrons in a column with a cross-sectional area of 1 $m^2$ along the path from the satellite to the receiver \cite{leitinger2008}. GPS data offer a highly effective means of estimating TEC values with improved spatial and temporal coverage \cite{ciraolo2007}. Given that the frequencies utilized in the GPS system are sufficiently high, the signals are only minimally influenced by ionospheric absorption and the Earth's magnetic field, both in the short-term and during long-term variations in the ionospheric structure \cite{nagarajoo2013}. Each Global Positioning System (GPS) satellite transmits signals at two distinct frequencies:
$f_1=1575.42$ MHz and
$f_2=1227.60 $ MHz. These signals use two different codes, P1 and P2, as well as two distinct carrier phases, L1 and L2, all derived from a common oscillator operating at 10.23 MHz \cite{gps2001}.\\
\vspace{0.3 cm}
In this study, the total electron content slant (STEC) derived from GPS data collected at our zero geographical latitude station in Galápagos islands is converted into the total vertical electron content (VTEC) as described in the literature \cite{hernandezpajares2002}:

\begin{equation}
STEC=\frac{1}{40.3}\left(\frac{f_1^2 \cdot f_2^2}{f_2^2-f_1^2}\right)\left[\left(P_1-P_2\right)+\left(\Delta b^s-\Delta b_r\right)\right], 
\end{equation}
where 
$b_ R$ and $b_S$
  represent the biases of the receiver and satellite, respectively.

\noindent
The vertical TEC is related to the slant TEC by:
\begin{equation}
    V T E C=  S(E_l)\cdot T E C, 
\end{equation}
where $S(E_l)$ is the translation coefficient, also known as the mapping function  at the ionospheric pierce point (IPP), and and it  is defined accordingly ( B. S. Arora, J. Morgan et al., 2015 \& references therein) as:

\begin{equation}
VTEC=\left(\sqrt{1-\left(\frac{R_T}{R_T+H} \cos(E_l)\right)^2}\right)\cdot STEC
\end{equation}

\noindent
$E_l$ indicates the satellite's elevation angle in degrees. VTEC reflects the vertical TEC at the IPP. $R_T$  denotes the Earth’s radius (6371 km) and for this study, we have selected an atmospheric height H of 400 km.\\

\vspace{0.3 cm}
The data for this analysis were obtained from GNSS stations in the CDDIS - NASA archive, focusing specifically on the total electron content (TEC) of the GLPS station during the period from May 1 to May 15, 2024. The RINEX data were filtered to include only satellites with an elevation angle above 40 degrees, ensuring a good signal-to-noise ratio. This data was supplemented with records from the magnetic data acquisition system (MAGDAS) sensor at the Quito Astronomical Observatory. Furthermore, the Kp and Dst indices, provided by the GFZ Helmholtz Center, were incorporated for the same date range (GFZ Helmholtz Center, 2024).\\
\vspace{0,3 cm}
Data from the MAGDAS station and TEC were processed using Fourier transform techniques and filtered to eliminate high-frequency noise. Data were organized into averaged blocks to align with the lengths of the Dst and Kp index data, facilitating correlation analysis \cite{press1992}.
Data were initially cleaned by removing outliers using the interquartile range (IQR) method, ensuring data accuracy. Subsequently, the data were grouped by time intervals and stations to calculate descriptive statistics like mean and standard deviation. To enhance the visualization and analysis of trends, spline interpolation was applied to smooth the curves.
Using a fast Fourier transform (FFT), the VTEC signal was decomposed into its constituent frequency components, revealing periodic patterns or cyclic variations. To enhance the visualization of long-term trends and eliminate high-frequency noise, a high-pass filter with a cutoff frequency of 0.00001 days was applied. This filtering effectively isolated the lower-frequency components, which are crucial for understanding the physical processes that govern the ionosphere, providing a more refined description of the underlying TEC trends.

 \section{G5 geomagnetic storm affectation on TEC}
 
We selected the extremely intense geomagnetic storm observed on 11 May 2024 for our study because of its exceptional strength, characterized by a minimum Dst-index of -410 nT and a Kp index of 9. Investigating the ionosphere's response to geomagnetic storms, particularly its impact on Total Electron Content (TEC), is a crucial aspect of space weather research. Normally, the ionosphere exhibits stable and predictable TEC levels, shaped by solar radiation, atmospheric dynamics, and seasonal factors. Under these typical conditions, ionization is predominantly driven by solar ultraviolet (UV) radiation, with electron densities peaking around local noon \cite{ratovsky2020}.

The intense storm of 11 May 2024 presents a valuable case for examining these processes in greater detail. By tracking the temporal evolution of TEC before, during, and after the storm, we aim to deepen our understanding of the ionosphere's behavior in response to such extreme geomagnetic disturbances. This research can contribute to improving predictive models of space weather impacts, which are crucial to mitigate the effects on satellite communications and navigation systems \cite{chaithra2024}.

During the main phase of a geomagnetic storm, negative storm effects often lead to substantial reductions in electron density, especially at low and equatorial latitudes. The reduced electron density weakens the ionospheric conductivity, thus diminishing current systems such as the equatorial electrojet (EEJ), which, in turn, leads to a reduction in the horizontal magnetic field \cite{simi2013}. The observed decrease of approximately 30 TEC units during the main phase of the storm can be attributed to a combination of factors, including enhanced recombination, plasma instabilities, atmospheric dynamics, and the disruption of normal ionospheric processes, all of which contribute to significant electron depletion \cite{fuller-rowell2008}

\section{Results and discussion}

\begin{figure}[h]
\centering
\includegraphics[scale=0.27]{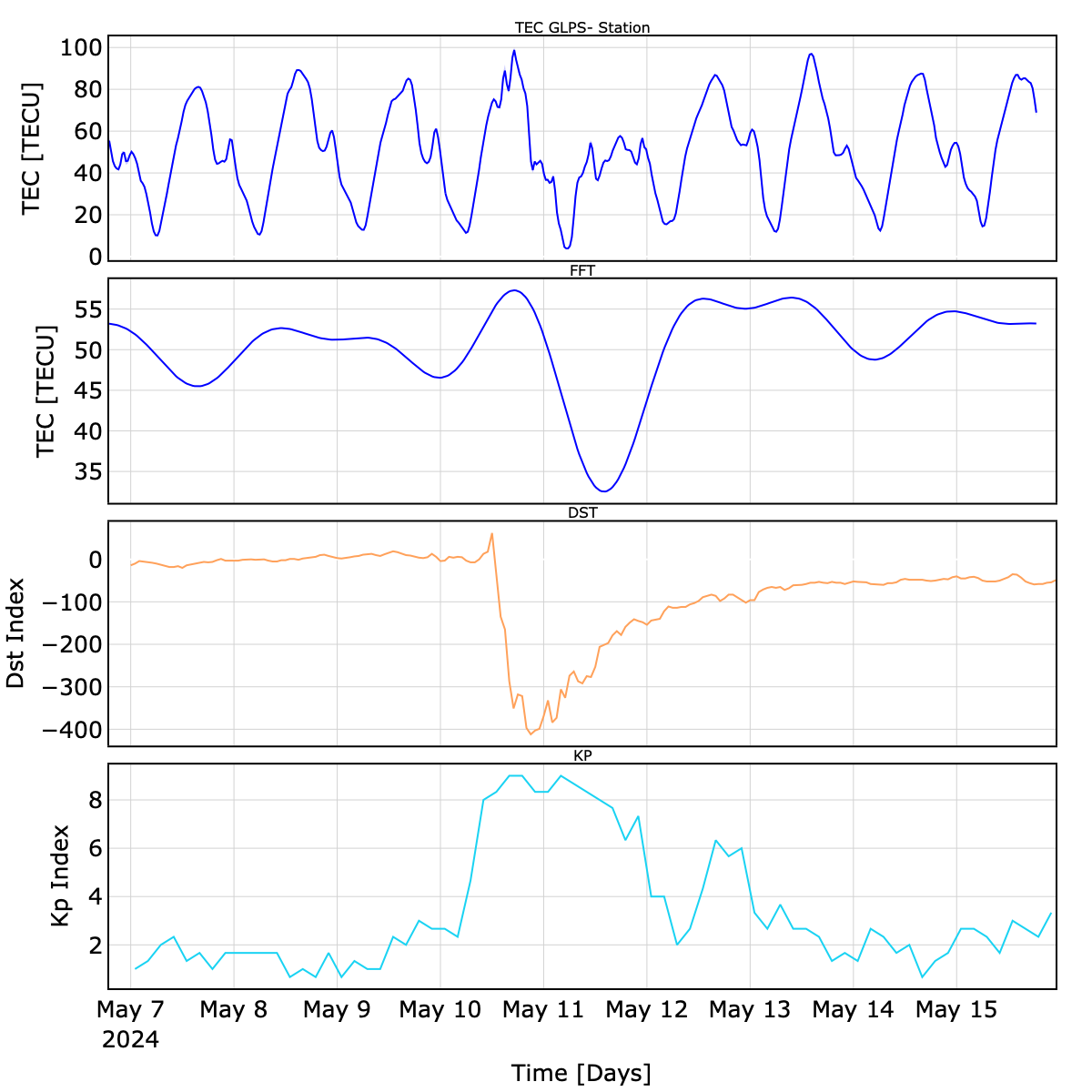}
\caption{Variation of Total Electron Content (TEC), Dst and Kp indexes, during the main phase of the may 2024, G5 storm}
\label{figure:1}
\end{figure}

\begin{figure}[h]
\centering
\includegraphics[scale=0.45]{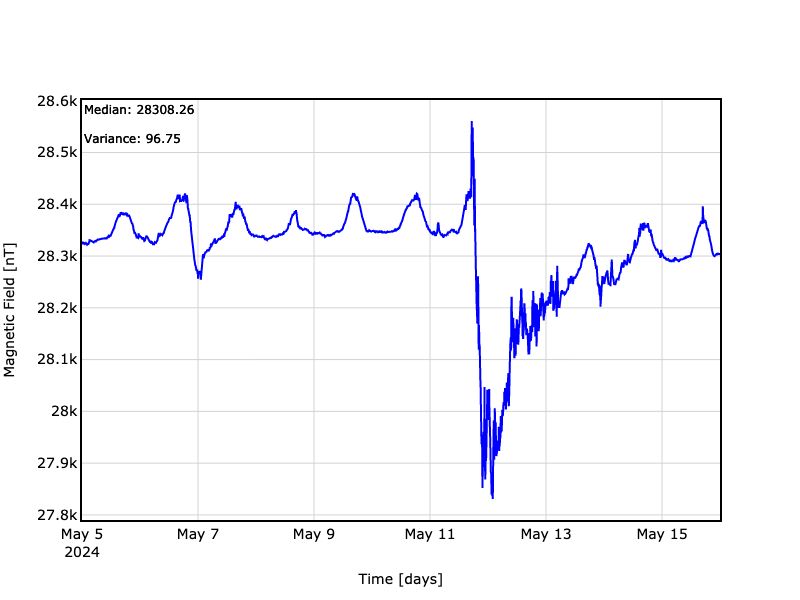}
\caption{May, 2024 G5 storm, horizontal magnetic field component}
\label{figure:2}
\end{figure}

 \begin{figure}[h]
\centering
\includegraphics[scale=0.45]{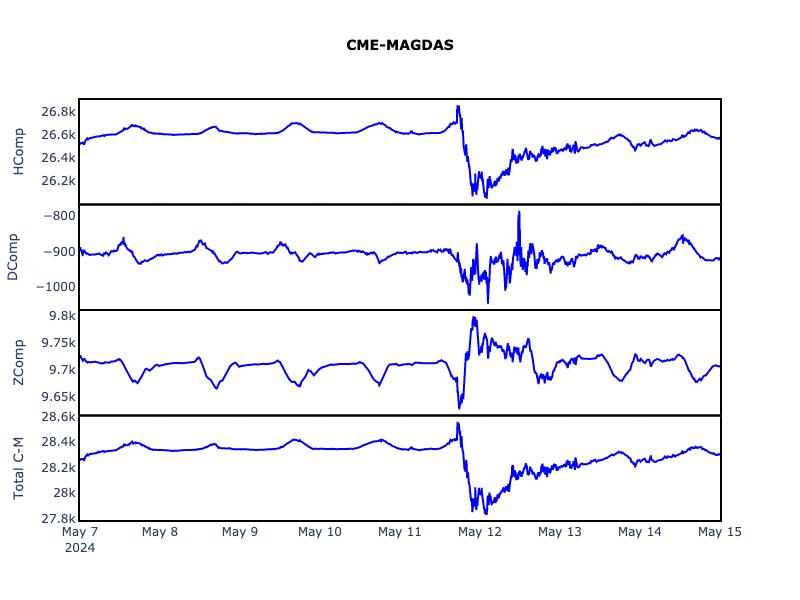}
\caption{May, 2024 G5 storm, magnetic field components}
\label{figure:2}
\end{figure}

 The analysis reveals a strong correlation between the CME (Coronal Mass Ejection) data and both the Dst and Kp indices, as illustrated in the accompanying graphs. This correlation suggests that the CME had a substantial impact on terrestrial geomagnetic indices, consistent with previous studies highlighting the significant influence of CMEs on space weather \cite{pulkkinen2017}. The global geomagnetic activity on 11 May 2024 reached a Kp index of 9, demonstrating the severity of the geomagnetic storm triggered by a massive CME and energetic solar wind. This event was classified as a G5 (extreme) storm, according to NOAA’s space weather scales.

The Earth's magnetic field disturbance, quantified by the Kp index, started increasing days before the CME's impact. The index rose from its normal level of Kp 2 on May 9, 2024, to a peak 9 on May 11, maintaining moderate activity until it returned to baseline on May 15.

The Disturbance Storm-Time (Dst) index, a measure of geomagnetic activity indicating global magnetic field perturbations, reached -410 nT on May 11, 2024. A Dst value this low signifies an exceptionally severe geomagnetic storm, known to weaken Earth's magnetic field due to intense solar activity, such as flares or CMEs \cite{kamide1998}. Geomagnetic storms of this magnitude pose serious risks to satellite operations, potentially disrupting or damaging satellite systems and communications. Moreover, intense geomagnetic storms can induce geomagnetically induced currents (GICs) in power grids, potentially damaging transformers and causing power outages \cite{pirjola2002}. While widespread outages were not reported during this storm, localized disruptions and voltage fluctuations occurred, particularly in areas vulnerable to GICs.

High-frequency (HF) radio communications, especially those used in aviation and maritime operations, experienced significant disruption during the storm, particularly in polar regions where geomagnetic disturbances were most severe \cite{lanzerotti2001}. The storm also produced a remarkable display of auroras, visible in regions far from the poles, including areas where such phenomena are rare. These auroras, observed globally, highlighted the extent of the solar wind's interaction with Earth's magnetosphere during this extreme event.

The geomagnetic storm persisted from May 10 to May 15, 2024, with the magnetosphere gradually returning to normal conditions. Reports from the ANEMOS/NKUA team indicate that this G5 extreme storm was caused by multiple CMEs that impacted Earth's magnetic field between May 10 and 13, intensifying the storm's duration and severity.

One notable effect of this storm was a significant decrease in the Total Electron Content (TEC), with a drop of nearly 30 units observed in the equatorial region. This reduction, contrary to the expected increase during such events, can be attributed to the unique ionospheric dynamics at low latitudes. In the equatorial ionosphere, processes like the equatorial anomaly and the electrojet are particularly sensitive to geomagnetic disturbances, leading to complex responses \cite{fuller-rowell2008}. Disruptions to the equatorial ionization anomaly (EIA) during the storm, caused by disturbance electric fields and atmospheric winds, likely contributed to the TEC reduction, particularly in the crest regions \cite{batista2011}.

Furthermore, the storm caused a significant reduction in the Earth's magnetic field strength, commonly referred to as "Dst depression." This decrease is primarily due to the enhanced ring current that forms during geomagnetic storms, as energetic particles from the solar wind are trapped in Earth's magnetosphere and circulate in the ring current. The magnetic field generated by these particles opposes Earth's own magnetic field, leading to the observed reduction, especially near the equator \cite{daglis1999}.

Typically, as a geomagnetic storm progresses, initial disturbances in Earth's magnetic field lead to sharp increases in ionization due to energetic particle precipitation. This can result in a short-term rise in TEC, followed by a gradual decline as the ionosphere returns to normal. The decay phase, during which TEC levels normalize, can take several hours to days, depending on the storm's intensity and duration. In some cases, these disturbances can cause long-lasting fluctuations in TEC, referred to as "storm-time ionospheric anomalies" \cite{fuller-rowell2008}.

\section{Conclusions}

We examined the effect of geomagnetic storms on the variation of TEC (Total Electron Content) at the low-latitude station in Galapagos, which lies just below the equatorial anomaly region. 
 
The geomagnetic storm on May 11, 2024, had a significant impact, decreasing TEC by approximately 30 TECU during the storm's main phase. During this phase, "negative storm effects" commonly lead to a substantial depletion of ionospheric electron density, especially at low latitudes. This depletion reduces the ionospheric conductivity and weakens ionospheric currents, which in turn lowers the strength of the horizontal magnetic field. The reduction in ionospheric conductivity during a geomagnetic storm, particularly at low latitudes, causes a noticeable decline in ionospheric currents like the equatorial electrojet (EEJ). The weakening of the EEJ directly contributes to the observed decrease in the horizontal component of the magnetic field. In equatorial regions, such a drastic reduction in the horizontal magnetic field is a clear indicator of the weakened ionospheric current systems.

Our analysis demonstrates a strong correlation between the coronal mass ejection (CME) that triggered this storm and the geomagnetic indices Dst and Kp, with the Dst index reaching -400 nT. This value signals a major geomagnetic storm with potentially serious consequences for Earth’s technological systems, underscoring the importance of continuous space weather monitoring to predict and mitigate such events. Monitoring CMEs, in particular, is crucial for forecasting geomagnetic disturbances.

Future research should aim to identify additional CME events and improve theoretical models describing magnetic field changes within CMEs. Moreover, real-time updates to ionospheric models, incorporating data from multiple sources, are essential for maintaining the accuracy of VTEC maps during geomagnetic storms. Advanced algorithms that can rapidly adjust to sudden changes in the ionosphere will be vital for enhancing space weather forecasting capabilities.

The May 11, 2024, geomagnetic storm serves as a reminder of the need for effective space weather monitoring and operational responses. While the storm had notable effects, they were largely mitigated due to preemptive monitoring and response efforts. This highlights the critical role of understanding and predicting space weather to safeguard technological infrastructure, including navigation and communication systems, from potential disruptions.

\section*{Data Availability}
The Global Navigation Satellite System (GNSS) data used in this study is publicly available from the NASA Center for Earth Observing System Data and Information Services (CEOS DIS) archive. Users can access daily GNSS data at: \url{https://cddis.nasa.gov/archive/gnss/data/daily}.

The calculated Vertical Total Electron Content (VTEC) data used in this study is not publicly available. However, interested researchers can request access to the VTEC data by contacting the corresponding author ([ericsson.lopez@epn.edu.ec]).

\section*{As Applicable – Inclusion in Global Research Statement}

\acknowledgments

%
%

\bibliography{referencias.bib}

%
%
%
%
%

\end{document}